\documentclass[%
 reprint,
 amsmath,amssymb,
 aps,
]{revtex4-1}

\usepackage{graphicx}
\usepackage{dcolumn}
\usepackage{bm}
\usepackage{hyperref}

\usepackage{color}

\begin{document}


\title{Collective effects in flow-driven cell migration}

\author{Louis Gonz\'alez}
\affiliation{Department of Physics and Astronomy, University of Pittsburgh, Pittsburgh, Pennsylvania 15260, USA}
\author{Andrew Mugler}
\email{andrew.mugler@pitt.edu}
\affiliation{Department of Physics and Astronomy, University of Pittsburgh, Pittsburgh, Pennsylvania 15260, USA}


\begin{abstract}
Autologous chemotaxis is the process in which cells secrete and detect molecules to determine the direction of fluid flow. Experiments and theory suggest that autologous chemotaxis fails at high cell densities because molecules from other cells interfere with a given cell's signal. Based on observations of collective cell migration in diverse biological contexts, we propose a mechanism for cells to avoid this failure by forming a collective sensory unit. Formulating a simple physical model of collective autologous chemotaxis, we find that a cluster of cells can outperform single cells in terms of the detected anisotropy of the signal. We validate our results with a Monte-Carlo-based motility simulation, demonstrating that clusters chemotax faster than individual cells. Our simulation couples spatial and temporal gradient sensing with cell-cell repulsion, suggesting that our proposed mechanism requires only known, ubiquitous cell capabilities. 
\end{abstract}

\maketitle


\section{Introduction}

One of the more fascinating ways that cells detect the direction of fluid flow is through a mechanism termed autologous chemotaxis. In autologous chemotaxis, cells secrete and bind to an autocrine factor that diffuses and drifts along the flow lines \cite{shields2007autologous}. More molecules bind to the downstream side of the cell, allowing it to determine the flow direction and consequently migrate downstream along the resulting concentration gradient \cite{shields2007autologous, fleury2006autologous}. Autologous chemotaxis is especially relevant in the context of metastatic cancer and has been observed in breast cancer cells \cite{shields2007autologous, polacheck2011interstitial}, melanoma cells \cite{shields2007autologous}, glioma cells \cite{munson2013interstitial}, as well as endothelial cells \cite{helm2005synergy}.

Experiments have found that autologous chemotaxis fails at high cell density and is overpowered by a competing, density-independent mechanosensing mechanism \cite{polacheck2011interstitial, polacheck2014mechanotransduction}. Theory \cite{Vennettilli2022} and simulations \cite{polacheck2011interstitial, Vennettilli2022} suggest that the reason for the failure is that, at high cell density, molecules secreted by other cells interfere with a given cell's autologous gradient. Essentially, the signal from all cells produces a background concentration which reduces the relative gradient experienced by any cell. A mean-field calculation based on this argument correctly predicts the cell density at which autologous chemotaxis fails \cite{Vennettilli2022}.

Nevertheless, in many other biological contexts, cells at high cell density have been shown to detect weak signals, including concentrations \cite{gregor2007probing} and concentration gradients \cite{rosoff2004new, ellison2016cell}. Theory has suggested that they do so by acting collectively \cite{erdmann2009role, Varennes:2017aa, mugler2016limits, camley2016collective, fancher2017fundamental}. Indeed, experiments have shown that collective sensing can lead to the detection of weaker signals \cite{ellison2016cell}, or to entirely different behaviors \cite{malet2015collective}, than cells can perform alone. These findings raise the question of whether autologous chemotaxis can benefit from these ubiquitous collective effects, to prevent sensory failure---or even mediate a sensory improvement---at high cell density. This question is particularly relevant to the dense tumor environment in which autologous chemotaxis is principally observed.

Here we combine theory, computational fluid mechanics, and Monte Carlo simulation to investigate the effects of collective sensing on autologous chemotaxis. We develop scaling arguments for how the detected signal should scale with cell density, and we validate these scalings by numerically solving the fluid flow and advection-diffusion equations describing the autocrine factor concentration. We then extend our results to dynamic cell migration simulations, revealing a regime in which cells chemotax faster as a cluster than as individuals. Our results reveal a novel chemotaxis mechanism based entirely on known and ubiquitous ingredients, with potential implications for migration of tumor cells and other cell types in high density environments.

\section{Results}
We first review in Sec.\ \ref{sec:individual} the results of our previous work on autologous chemosensing by an individual cell that will be useful in generalizing to collective autologous chemosensing. Then, in Sec.\ \ref{sec:scalingargument} we derive how the strength of collective autologous chemosensing should scale with the cell density. In Sec.\ \ref{sec:comsolsimulation}, we compare our theoretical results to the numerical solution of the fluid mechanics problem and identify a crossover density at which collective sensing outperforms individual sensing. Lastly, in Sec.\ \ref{sec:montecarlosimulation}, we demonstrate using a motility simulation that better sensing by a cell cluster results in faster migration velocity along the flow direction.

\subsection{Individual autologous chemosensing}
\label{sec:individual}

Information about the flow direction is contained in the imbalance between the numbers of molecules detected upstream versus downstream. For an individual cell, this imbalance is quantified using the anisotropy measure \cite{fancher2020precision, endres2008accuracy, Varennes:2017aa}
\begin{equation}
\label{AIsum}
A_I =\frac{1}{M}\sum_{i=1}^M\cos\theta_i.
\end{equation}
Here, $\theta_i$ is the angle, relative to the flow direction, of the return of the $i$th molecule to the surface of the cell, out of $M$ total returning molecules. The cosine extracts the asymmetry between the downstream ($\theta = 0$) and upstream ($\theta = \pi$) sides of the cell such that $A_I > 0$ for a downstream gradient and $A_I < 0$ for an upstream gradient.  In this work we assume that a cell passively detects molecules, e.g., by receptor binding and unbinding, rather than permanently absorbing them, although we discuss absorption elsewhere \cite{fancher2020precision, Vennettilli2022}.

In previous work \cite{Vennettilli2022}, we showed that the anisotropy could be approximated as $A_I \approx (n_d-n_u)/n_d$, where $n_u$ and $n_d$ are the numbers of molecules detected by the upstream and downstream halves of the cell, respectively. Specifically,
\begin{equation}
\label{nup}
n_u = \frac{\nu}{D/a^2+v/a},
\end{equation}
where $\nu$ is the molecule secretion rate, $D$ is the molecular diffusion coefficient, $a$ is the radius of the cell, and $v$ is the flow speed. Eq.\ \ref{nup} constructs the molecule number as a ratio of the rates of molecules entering (by secretion) and leaving (by diffusion or flow, respectively) the cell half. The expression for $n_d$ lacks the $v_0/a$ term because molecules lost to flow downstream are replenished by those lost to flow from the upstream half. As a result, the anisotropy simplifies to $A_I \approx \epsilon$ for small P\'eclet number $\epsilon = v_0a/D$ \cite{Vennettilli2022} (experiments suggest that indeed $\epsilon \ll 1$ for autologous chemotaxis \cite{shields2007autologous, polacheck2011interstitial, fancher2020precision}). A more rigorous calculation confirms this scaling, yielding $A_I = \epsilon/8$ \cite{fancher2020precision}.

Autologous chemotaxis for an individual cell fails at high cell density. Specifically, in previous work \cite{Vennettilli2022} we showed using a mean-field argument that, in the presence of identical cells at a density $\rho$, the anisotropy for a given cell scales as
\begin{equation}
\label{AI}
A_I = \frac{\epsilon/8}{1 + \rho/\rho_c},
\end{equation}
where $\rho_c = \epsilon/4\pi a^2 L$, and $L$ is the system size in the flow direction. The critical density $\rho_c$ is the cell density beyond which sensing begins to fail due to the presence of molecules secreted by other cells.

\subsection{Scaling argument for collective autologous chemosensing}
\label{sec:scalingargument}

\begin{figure}
    \centering
    \includegraphics[width = \columnwidth]{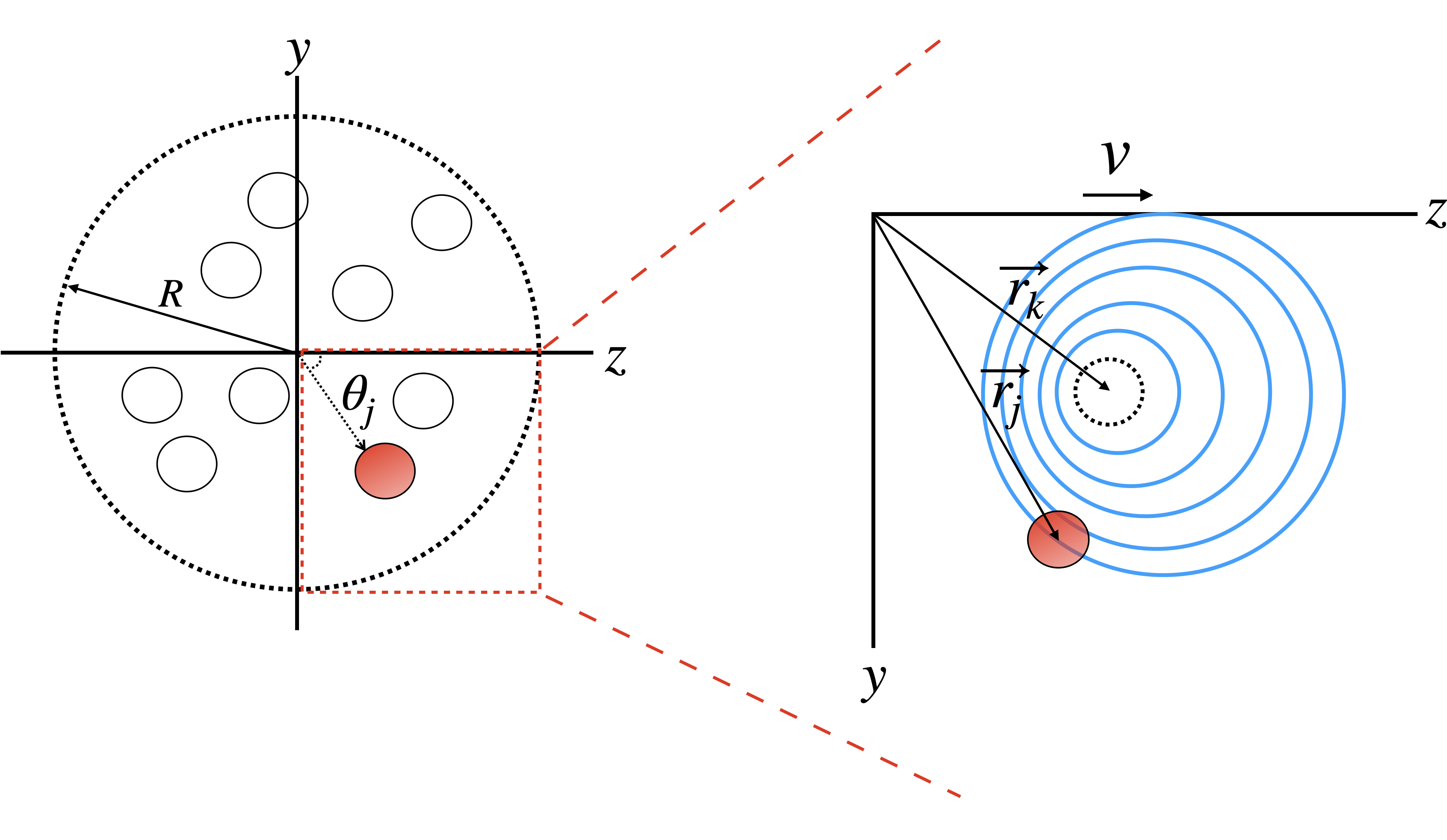}
    \caption{Schematic of the cell collective. Left: a particular cell $j$ makes an angle $\theta_j$ with the flow direction $\hat{z}$, relative to the collective's center of mass. The collective's volume is characterized by a lengthscale $R$. Right: Molecules secreted by any other cell $k$ drift in the flow direction a distance $vt$ and diffuse isotropically a characteristic distance $\sqrt{Dt}$, tracing out a spherical shell.}
    \label{fig:autologous3sketch}
\end{figure}

We now consider an entire collective of cells as the sensory unit (Fig. \ref{fig:autologous3sketch}). Collective sensing has been investigated in the context of external gradient detection \cite{mugler2016limits, camley2016collective} and has been suggested to account for the ability of groups of cells to detect shallower gradients than any individual cell \cite{ellison2016cell}. Here we investigate whether collective sensing can rescue, or even improve, autologous chemotaxis at high cell density.

Collective anisotropy is defined similarly to Eq. \ref{AIsum}, with the key difference that $N$ cells now perform a sensory calculation as a single unit (we will elaborate on how they could do so in Sec.\ \ref{sec:montecarlosimulation}). Specifically,
\begin{equation}
\label{AC}
A_C = \frac{1}{n_T} \sum_{j=1}^N n_j \cos \theta_j.
\end{equation}
Here, $n_j$ is the number of molecules in the vicinity of the $j$th cell, $\theta_j$ is the angle the cell makes with the flow direction (relative to the collective's center of mass), and $n_T$ is the total number of molecules in the vicinity of the collective (Fig. \ref{fig:autologous3sketch}, left). We estimate $n_j$ and $n_T$ following Eq.\ \ref{nup},
\begin{align}
\label{nj}
n_j &= \frac{\nu + \nu_j}{D/a^2 + v/a}, \\
\label{nT}
n_T &= \frac{N\nu}{D/R^2+v/R}.
\end{align}
In Eq.\ \ref{nj}, $\nu_j$ is the rate of arrival, to cell $j$, of molecules secreted by other cells. In Eq.\ \ref{nT}, $R$ is the radius of the collective; for a spherical arrangement, it is related to the cell density as $\rho = N/(4\pi R^3/3)$.

To find the molecule arrival rate $\nu_j$, we consider a specific cell $k$ in the collective that acts as a source of these molecules, and we will ultimately sum over $k$. In a time $t$, a molecule released from cell $k$ drifts in the flow direction a distance $vt$ and diffuses isotropically a characteristic distance $\sqrt{Dt}$, tracing out a spherical shell described by $|\vec{r}-\vec{r}_k-vt\hat{z}|^2 = Dt,$
where $\hat{z}$ is the flow direction, and $\vec{r}_k$ is the position of cell $k$ (Fig.\ \ref{fig:autologous3sketch}, right). This shell will reach cell $j$ when $\vec{r}=\vec{r}_j$, giving $|\vec{r}_j-\vec{r}_k|^2-2vt(z_j-z_k)+v^2t^2 = Dt$. Rescaling time as $\tau\equiv tD/a^2$ and recalling that $\epsilon = va/D$, this equation becomes $|\vec{r}_j-\vec{r}_k|^2/a^2-2\epsilon\tau(z_j-z_k)/a+\epsilon^2\tau^2 = \tau$. Because the P\'eclet number is small ($\epsilon\ll1$), we neglect the quadratic term, giving a rescaled arrival time of $\tau = |\vec{r}_j-\vec{r}_k|^2/[a^2+2\epsilon a(z_j-z_k)]$. At this time, the shell has a radius $\sqrt{Dt}$, and the likelihood of the molecule reaching cell $j$ is the ratio of the cell's cross-sectional area $\pi a^2$ to the shell's surface area $4\pi Dt$, or $a^2/4Dt = 1/4\tau$. Thus, the arrival rate of molecules at cell $j$ is the secretion rate $\nu$ multiplied by this likelihood and summed over $k$,
\begin{equation}
\label{nuj}
\nu_j = \nu\sum_{k\ne j} \frac{a^2+2\epsilon a(z_j-z_k)}{4|\vec{r}_j-\vec{r}_k|^2}.
\end{equation}

We insert Eq.\ \ref{nuj} into Eq.\ \ref{nj}, and Eqs.\ \ref{nj} and \ref{nT} into Eq.\ \ref{AC}. For the purposes of obtaining a scaling, we approximate the sums as integrals. Doing so, and writing Eqs.\ \ref{nj} and \ref{nT} in terms of $\epsilon$, we obtain
\begin{align}
\label{ACint}
A_C \approx\ &\frac{a^2[1+(R/a)\epsilon]}{NR^2(1+\epsilon)}\int\frac{d^3r_j}{R^3/N} \cos\theta_j \Bigg[1+ \nonumber \\
	&\int\frac{d^3r_k}{R^3/N}
	\frac{a^2+2\epsilon a(r_j\cos\theta_j-r_k\cos\theta_k)}{4|\vec{r}_j-\vec{r}_k|^2}\Bigg].
\end{align}
Here we have used $z=r\cos\theta$ and scaled the volume element $d^3r$ by the typical volume occupied by one cell, which goes as $R^3/N$. Within the large square brackets in Eq.\ \ref{ACint}, any term that does not depend on $\theta_j$ will vanish by symmetry when integrated against the $\cos\theta_j$ outside. Therefore, we isolate the middle term of the $\vec{r}_k$ integral \footnote{Neglecting the first and last terms in the second line of Eq.\ \ref{ACint} ignores the dependence of $|\vec{r}_j-\vec{r}_k|^2$ on $\theta_j$. We validate this uncontrolled approximation post hoc when checking our results against numerics in Sec.\ \ref{sec:comsolsimulation}.},
\begin{equation}
\label{ACint2}
A_C \approx\ \frac{\epsilon a^3N[1+(R/a)\epsilon]}{2R^8(1+\epsilon)}
	\int d^3r_jd^3r_k\frac{r_j\cos^2\theta_j}{|\vec{r}_j-\vec{r}_k|^2}.
\end{equation}
In the prefactor of Eq.\ \ref{ACint2}, we may neglect the additive terms proportional to $\epsilon$ as long as $R/a$ is not too large. In the integral in Eq.\ \ref{ACint2}, we can understand how the result should scale with $R$ without performing the integration: the volume elements contribute factors of $R^3$ each because the integration extends out to $r=R$; and the numerator and denominator contribute factors of $R$ and $R^{-2}$, respectively. Altogether, we have $A_C \sim \epsilon a^3N/R^3$, or
\begin{equation}
\label{ACscale}
A_C \sim \epsilon a^3\rho,
\end{equation}
where we have recognized $\rho \sim N/R^3$ as the cell density.

Equation\ \ref{ACscale} is our main result for how the collective anisotropy should scale with system properties. Several features make intuitive sense. First, the collective anisotropy should vanish as the cell density $\rho$ gets small. The reason is that when cells are far apart, each cell detects the same number of molecules (its own), and the collective computation yields no information on upstream-downstream molecule imbalance. Second, the collective anisotropy should increase with the P\'eclet number $\epsilon = va/D$, as the individual anisotropy does (Eq.\ \ref{AI}). The reason is that a larger $\epsilon$ (e.g., via a faster flow speed $v$) naturally increases the molecule imbalance.

\subsection{Numerical validation and crossover cell density}
\label{sec:comsolsimulation}

Here we validate Eq.\ \ref{ACscale} by numerically solving the fluid dynamics and advection-diffusion equations. Specifically, we solve the steady-state Brinkman equation (appropriate for the low-Reyolds-number, low-permeability cell environment \cite{polacheck2011interstitial, shields2007autologous}) to find the velocity field. The velocity field provides the advection term in the steady-state advection-diffusion equation for the molecular concentration. We use a finite-element computational fluid dynamics package (COMSOL) to solve both equations \cite{Vennettilli2022, polacheck2011interstitial}. Details are provided in our previous work \cite{Vennettilli2022} and the code for the present work is freely available \cite{code}.

\begin{figure}
    \centering
    \includegraphics[width = \columnwidth]{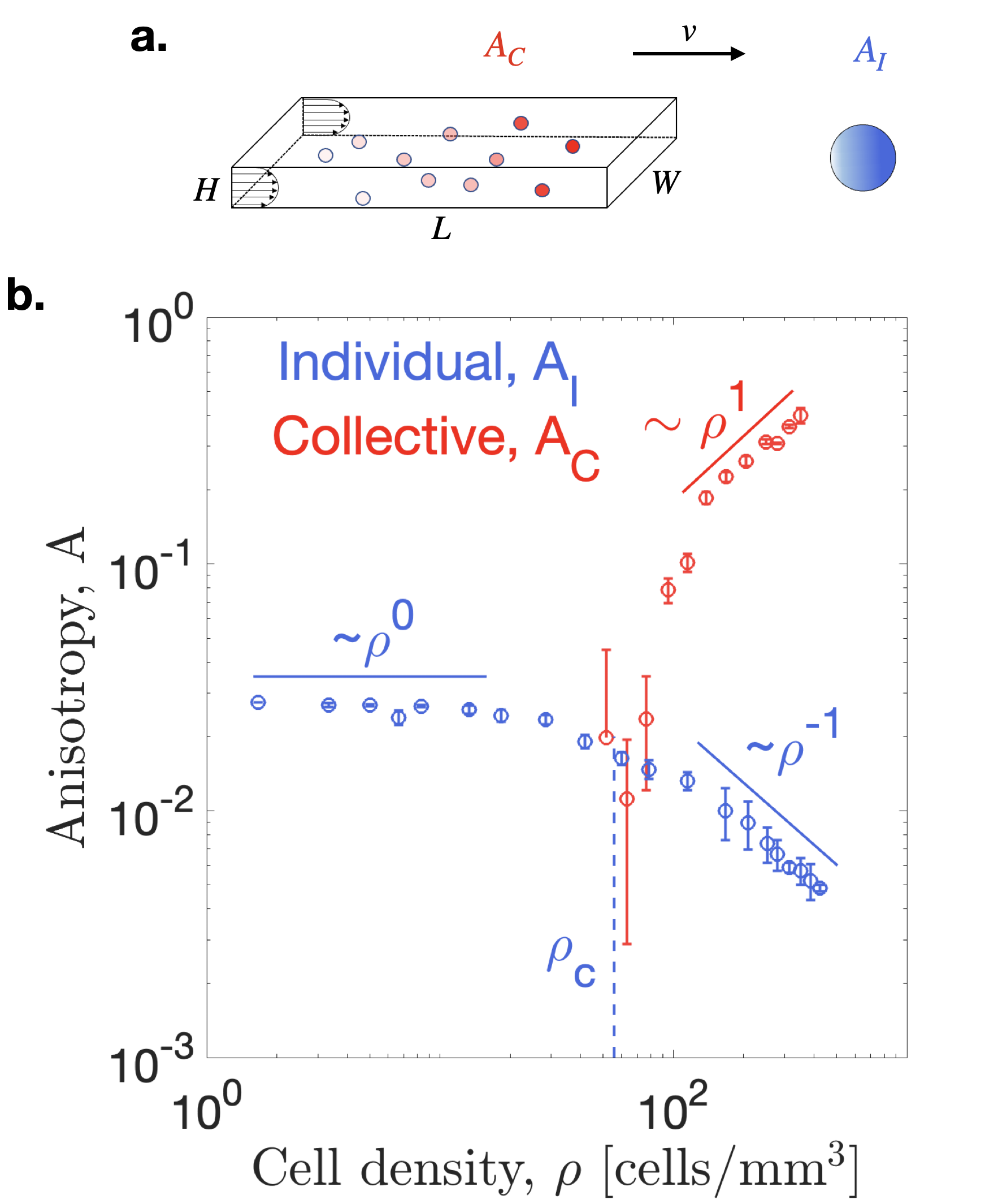}
    \caption{Numerical computation using fluid dynamics software. (a) Schematic of cells randomly placed in domain with dimensions $L$, $W$, and $H$, and flow in the $L$ direction. Collective anisotropy $A_C$ considers the average surface concentration at each cell (left), whereas individual anisotropy $A_I$ considers the angular variation in the surface concentration around the center cell (right). (b) Individual and collective anisotropy vs.\ cell density from numerics (mean and standard error over 5 trials with random cell configurations), compared with predicted scalings. Here $a = 10 \,\mu \text{m}$, $\nu = 1$ s$^{-1}$, $D = 150 \, \mu \text{m}^2/\text{s}$, $v = 3 \, \mu \text{m}/\text{s}$, $K = 0.1 \, \mu \text{m}^2$, $L = 3 \, \text{mm}$, $W \approx 2 \, \text{mm}$, and $H \sim 100 \, \mu \text{m}$.}
    \label{fig:autologous3fig1}
\end{figure}

To vary the cell density, we keep the volume of the system constant while varying the number of cells $N$. This protocol mimics the microfluidic experiments used to investigate autologous chemotaxis \cite{polacheck2011interstitial}, and indeed we consider a system with dimensions similar to the microfluidic chamber: a rectangular box domain of length $L$, width $W$, and height $H$, where the flow is in the direction of $L$ (Fig.\ \ref{fig:autologous3fig1}a). Cells are placed uniform-randomly throughout the box, ensuring that one cell is in the center, and that the cells do not overlap with one another or with the boundaries of the box. Both the individual and collective anisotropy measures are averaged over random configurations of cells in the domain.

The individual anisotropy $A_I$ follows from Eq.\ \ref{AIsum}, which, in terms of the continuous molecular concentration at the cell surface $c(a,\theta,\phi)$, is equivalent to \cite{fancher2020precision}
\begin{equation}
\label{eqn:anisotropy}
	A_I = \frac{\int d\Omega c(a,\theta,\phi) \cos\theta}{\int d\Omega' c(a,\theta',\phi')},
\end{equation}
where $d\Omega = d\phi d\theta \sin\theta$ is the solid angle element. We compute these integrals from the numerical solution for $c$ at the surface of the center cell \cite{Vennettilli2022} (Fig.\ \ref{fig:autologous3fig1}a, right).

The collective anisotropy $A_C$ follows from Eq.\ \ref{AC}, which, in terms of the surface concentration $c_j$ averaged around cell $j$, reads
\begin{equation}
A_C = \frac{\sum_{j=1}^N c_j \cos \theta_j}{\sum_{k=1}^N c_k}.
\end{equation}
We compute $c_j$ from the numerical solution at the surface of each cell (Fig.\ \ref{fig:autologous3fig1}a, left).

The model parameters are set from experiments. A breast cancer (MDA-MB-231) cell is approximately $a = 10 \,\mu \text{m}$ in radius \cite{shields2007autologous, polacheck2011interstitial} and secretes approximately $\nu = 1$ CCL19/21 molecule per second \cite{shields2007autologous, fancher2020precision} which diffuses with approximate coefficient $D = 150 \, \mu \text{m}^2/\text{s}$ \cite{fleury2006autologous}. The cell density experiments \cite{polacheck2011interstitial} were performed with flow velocity $v = 3 \, \mu \text{m}/\text{s}$ and permeability $K = 0.1 \, \mu \text{m}^2$ in a chamber of length $L = 3 \, \text{mm}$, width $W \approx 2 \, \text{mm}$, and height $H \sim 100 \, \mu \text{m}$.

The numerical anisotropies as a function of cell density $\rho = N/LWH$ are shown in Fig.\ \ref{fig:autologous3fig1}b. We expect from Eq.\ \ref{AI} that the individual anisotropy $A_I$ should scale as $\rho^0$ for $\rho \ll \rho_c$ and as $\rho^{-1}$ for $\rho \gg \rho_c$, and we see in Fig.\ \ref{fig:autologous3fig1}b (blue) that the numerics agree, as seen previously \cite{Vennettilli2022}. We expect from Eq.\ \ref{ACscale} that $A_C$ should scale with $\rho$, and we see in Fig.\ \ref{fig:autologous3fig1}b (red) that the numerics agree at large $\rho$.

We also expect the scaling in Eq.\ \ref{ACscale} to break down when the typical distance between cells $\rho^{-1/3}$ becomes larger than the smallest lengthscale of the domain (here, $H$). We write this condition as $\rho^{-1/3} > \alpha H$, where $\alpha$ is a constant that we expect to be of order unity. Rearranging, we have $\rho < (\alpha H)^{-3}$. Fig \ref{fig:autologous3fig1}b (red) shows that the numerics indeed become especially sensitive to cell arrangement, leading to large variability in $A_C$, for $\rho \lesssim 10^2$ mm$^{-3}$, corresponding to $\alpha \approx 2.1$, which is indeed of order unity.

Figure \ref{fig:autologous3fig1}b demonstrates that collective sensing outperforms individual sensing ($A_C>A_I$) above a crossover density on the order of $\rho \sim 50$ cells/mm$^{-3}$. Indeed, autologous chemotaxis has been observed in the range $50$$-$$250$ cell/mm$^{-3}$ \cite{polacheck2011interstitial}, and individual sensing is thought to break down toward the top of that range \cite{polacheck2011interstitial, Vennettilli2022}. The typical cell spacing at the crossover density, $\rho^{-1/3}\sim270$ $\mu$m, is much larger than a cell diameter, $2a\approx 20$ $\mu$m, implying that collective effects could be beneficial well before reaching the tight-packing limit typical of tissues and tumors.

\subsection{Motility simulation and collective chemotaxis}
\label{sec:montecarlosimulation}

The previous section demonstrated that beyond a crossover density, cells sense the flow direction better collectively than individually. Presumably better chemosensing leads to faster chemotaxis, but this hypothesis must be checked. Moreover, it is not clear how the information sharing required in our definition of collective anisotropy is achievable by individual cells. To these ends, here we develop a motility simulation and measure the chemotaxis speed explicitly. Our simulation incorporates only concentration sensing and gradient sensing by individual cells, and cell-cell repulsion. We will see that these basic capabilities, ubiquitous among cells of many types, are sufficient to realize the density-mediated crossover from individual to collective chemotaxis in the flow direction.

To focus on the basic physics and maintain computational tractability, the simulation makes two important simplifications. First, we reduce cells to point particles on a cubic lattice with spacing given by the cell radius $a$. Thus, each cell moves to one of six neighboring sites at each time step according to a Monte Carlo scheme, as described shortly. Second, we write the molecular concentration field as the sum of contributions from each cell, where each contribution is approximated as the known single-cell solution obtained as if the cell were isolated \cite{fancher2020precision}. This approximation avoids the need to numerically solve for the flow lines and the concentration field at every time step, and it should be valid for cell densities not too close to the tight-packing limit.

Specifically, we approximate the concentration as $c(\vec{r}) \approx \sum_{j=1}^N \tilde{c}(\vec{r}-\vec{r}_j)$, where $\vec{r}_j$ is the position of cell $j$, and $\tilde{c}$ is the steady-state solution to the single-cell problem, solved previously \cite{fancher2020precision} using the P\'eclet number $\epsilon = va/D$ as a perturbation parameter. That solution is
\begin{equation}
\frac{\tilde{c}(\vec{r})}{\bar{c}} = \frac{a}{r}+\frac{\epsilon}{2}\left\{-1
	+\frac{\cos\theta}{4}\left[\frac{f(1)}{2r^2/a^2}+f(r/a)\right]\right\},
\end{equation}
where $\bar{c} = \nu/4\pi Da$,
\begin{align}
    &f(x) = 4 - \frac{4(2\kappa+1)}{x^2} + \frac{2(1+3\kappa+3\kappa^2)}{x^3} \nonumber \\
    &+ \frac{\kappa^2e^{1/\kappa}}{x^3}\left[\left( \frac{x^3}{\kappa^3} - \frac{x^2}{\kappa^2} +
    \frac{2x}{\kappa} - 6\right)e^{-x/\kappa} - \frac{x^4E_1(x/\kappa)}{\kappa^4}\right],
\end{align}
$E_1(y) = \int_1^\infty dt\ e^{-yt}/t$, and $\kappa = \sqrt{K}/a$ for permeability $K$. We note that $f(1)$ varies between $1$ ($\kappa\gg1$) and $2$ ($\kappa\ll1$). For these simulations we keep $a = 10 \,\mu \text{m}$, $\nu = 1$ s$^{-1}$, $D = 150 \, \mu \text{m}^2/\text{s}$, and $K = 0.1 \, \mu \text{m}^2$ as above, but we lower the flow speed to $v = 0.5 \, \mu \text{m}/\text{s}$, thus lowering the P\'eclet number to $\epsilon = 1/30$, in order to maintain the validity of the perturbative solution out to distances much larger than the cell size.

The Monte Carlo scheme accepts or rejects moves according to a potential energy and a work function \cite{szabo2010collective, varennes2016collective, roy2021intermediate}. Concentration sensing is incorporated into the potential energy. The energy difference involved in the Monte Carlo scheme is then equivalent to comparing concentration values from one time step to the next, akin to temporal gradient sensing, as seen in motile bacteria \cite{macnab1972gradient, mao2003sensitive}. Gradient sensing is incorporated into the work function. Computing the work is then equivalent to comparing concentration values at neighboring lattice points in a single time step, akin to spatial gradient sensing, as seen in amoeba and yeast \cite{arkowitz1999responding, swanson1982local}. Cell-cell repulsion is incorporated into the potential energy. Cell-cell repulsion occurs in many cell types, often mediated by contact inhibition of locomotion \cite{mayor2010keeping}.

The potential energy is then
\begin{equation}
U = \sum_{j=1}^N\sum_{k<j} \frac{\lambda^2}{|\vec{r}_j-\vec{r}_k|^2} - \psi_c\sum_{j=1}^N \frac{c(\vec{r}_j)}{\bar{c}}.
\end{equation}
In the first term, closer cell pairs correspond to larger energy. This term thus corresponds to cell-cell repulsion, with length parameter $\lambda$. In the second term, larger concentration values correspond to smaller energy. This term thus corresponds to concentration sensing with strength $\psi_c$. Singularities in the second term from self-energies $\tilde{c}(0)$ are removed because the Monte Carlo scheme considers only energy differences between configurations before and after a cell moves, which contain the same $N$ self-energy terms.

The work function is
\begin{equation}
\label{work}
W = \psi_g\frac{c(\vec{r}_j+\delta\vec{r}_i) - \bar{c}_j}{\bar{c}_j},
\end{equation}
where $\delta\vec{r}_i/a$ are the unit vectors in each of the six directions, and $\bar{c}_j = \sum_{i=1}^6 c(\vec{r}_j+\delta\vec{r}_i)/6$ is the concentration averaged over these neighboring sites. Equation \ref{work} gives the work corresponding to the movement of cell $j$ to its neighboring site in direction $i$. Positive work means moving to a site whose concentration is higher than the average of all neighboring sites. Equation \ref{work} thus corresponds to gradient sensing with strength $\psi_g$.

Given the energy and work terms, the Monte Carlo scheme proceeds as follows \cite{varennes2016collective, roy2021intermediate}. At each time step, each cell $j$ moves to its neighboring site $i$ (selected at random) with probability
\begin{equation}
P = \begin{cases}
e^{-(\Delta U - W)} & \Delta U - W \ge 0 \\
1 & \Delta U - W < 0,
\end{cases}
\end{equation}
where $W$ is calculated before the move, and $U$ is calculated both after and before the move to give $\Delta U$. The cells are initialized as an $N$-cell chain along the direction of the flow and move in an unbounded domain.

At low cell density, $\Delta U\to0$, and $W$ for a move in the flow direction is on the order of $\psi_g\epsilon$. Therefore we set $\psi_g$ to a value on the order of $1/\epsilon$, namely $\psi_g = 10$. We then vary the relative strength of concentration sensing vs.\ gradient sensing by varying $\psi_c$. To do so in a way that maintains a typical spacing between cells, we consider the potential energy between a pair of cells separated by a distance $r$, which reads $U = \lambda^2/r^2 - 2\psi_c[a/r+\tilde{c}(0)/\bar{c} + \mathcal{O}(\epsilon)]$. This function has a minimum at $r_* = \lambda^2/\psi_ca$. Therefore, for a given $r_*$, as we vary $\psi_c$, we set $\lambda$ via this expression until it becomes too small to mediate the repulsion. Specifically, we find that $\lambda = \max(\sqrt{\psi_car_*},10a)$ is sufficient to prevent cells from cohering permanently (which arrests migration). 

\begin{figure*}
    \centering
    \includegraphics[width = 6in]{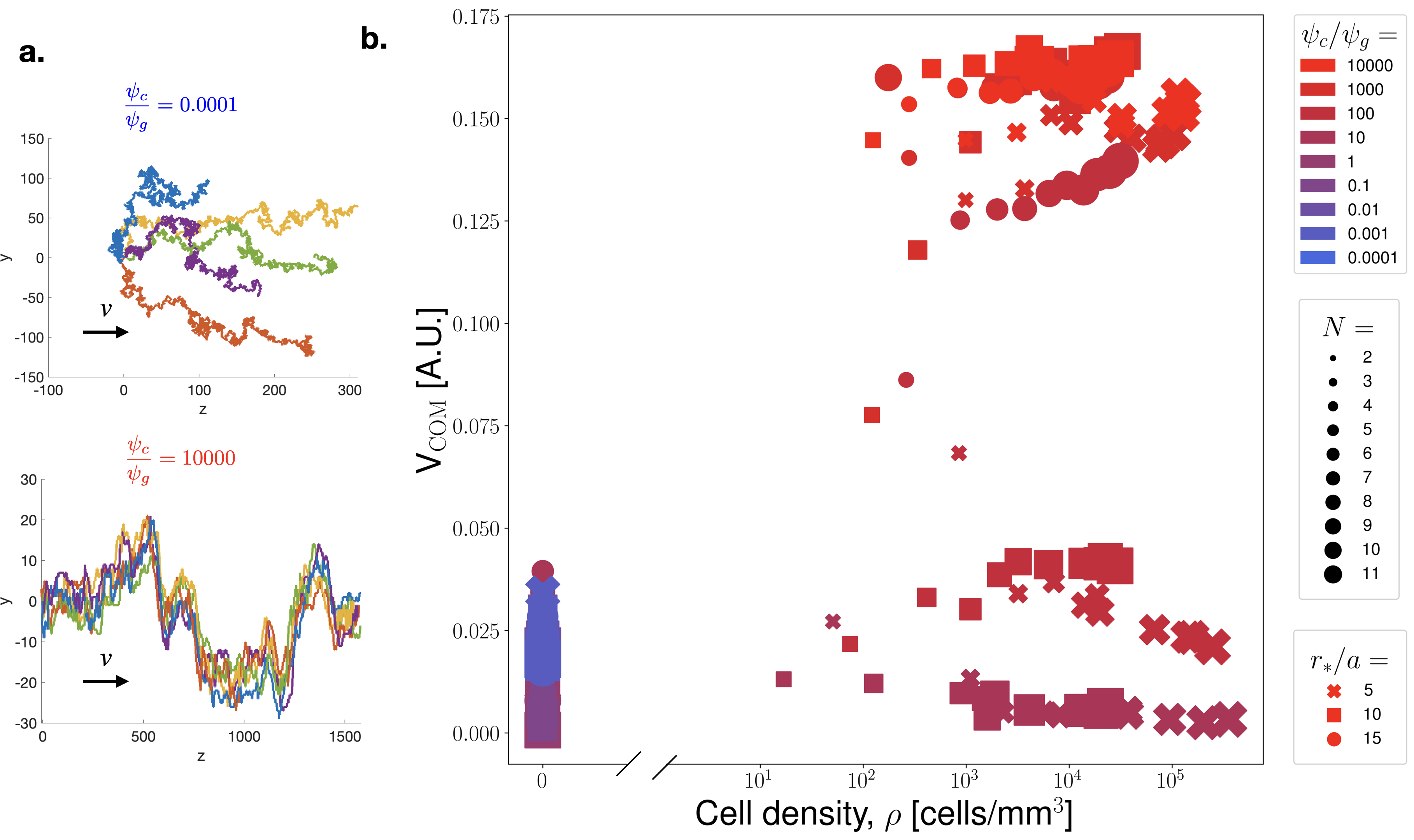}
        \caption{Motility simulation. (a) Snapshots of $N=5$ cell trajectories with $r_*/a = 1$, for $\psi_c\ll\psi_g$ (top) and $\psi_c\gg\psi_g$ (bottom). Cells drift in the flow direction individually (top) or collectively (bottom). Note the difference in axes' scales: the migration distance is much farther for the bottom plot. (b) Center-of-mass velocity vs.\ cell density. As $\psi_c/\psi_g$ increases (blue to red), density increases as cells transition to collective migration. The velocity first decreases, then increases, indicating that collective chemotaxis outperforms individual chemotaxis. Here $a = 10 \,\mu \text{m}$, $\nu = 1$ s$^{-1}$, $D = 150 \, \mu \text{m}^2/\text{s}$, $v = 0.5 \, \mu \text{m}/\text{s}$, $K = 0.1 \, \mu \text{m}^2$, $\psi_g = 10$, and $\lambda = \max(\sqrt{\psi_car_*},10a)$.}
\label{fig:autologous3fig2}
\end{figure*}

Figure \ref{fig:autologous3fig2} shows the simulation results. Focusing first on typical cell trajectories (Fig.\ \ref{fig:autologous3fig2}a), we see that for $\psi_c\ll \psi_g$ (top), cells execute diffusive trajectories that drift in the flow direction but do not stay together. This makes sense: at low $\psi_c$, cells lack the coattraction mediated by concentration sensing and feel only repulsion when close; once separated, they execute autologous chemotaxis individually by spatial gradient sensing. In contrast, we see that for $\psi_c\gg\psi_g$ (bottom), cells remain as a cohesive group whose center of mass executes a diffusive trajectory that drifts in the flow direction. This also makes sense: at high $\psi_c$, concentration sensing mediates both a coattraction and the movement toward maximal concentration; due to the flow, the highest concentration is downstream of the group, resulting in collective autologous chemotaxis.

In Fig.\ \ref{fig:autologous3fig2}b, we plot the center-of-mass velocity in the flow direction $v_{\rm COM}$ vs.\ the cell density $\rho$, as we vary $\psi_c/\psi_g$ (color), $r_*$ (shape), and $N$ (size). The center-of-mass velocity is computed as the displacement in the flow direction divided by the number of time steps, for $10^4$ time steps, averaged across the $N$ cells and across five simulation trials. The cell density is computed as the inverse of the average cell-cell spacing. In simulations for which the average cell-cell spacing does not saturate within $10^4$ time steps, cells are determined to be diffusing away from each other indefinitely, and the cell density is set to $\rho = 0$.

Consistent with Fig.\ \ref{fig:autologous3fig2}a, we see in Fig.\ \ref{fig:autologous3fig2}b that the results are relatively insensitive to $r_*$ and $N$, and are primarily tuned by $\psi_c/\psi_g$. Specifically, we see that for $\psi_c\ll\psi_g$ (blue, lower left), $\rho = 0$, and cells move with a characteristic $v_{\rm COM}$ indicative of individual autologous chemotaxis. For $\psi_c \sim \psi_g$ (purple, lower right), $\rho$ increases as the coattraction sets in, and $v_{\rm COM}$ slightly decreases. Interestingly, this finding is consistent with the observation that increased cell density reduces the individual anisotropy (Fig.\ \ref{fig:autologous3fig1}b, lower right). Finally, for $\psi_c\gg\psi_g$ (red, upper right), $\rho$ stays high, and $v_{\rm COM}$ significantly increases, above that for $\psi_c\ll\psi_g$. This finding indicates that collective chemotaxis can outperform individual chemotaxis, also consistent with the observation in Fig.\ \ref{fig:autologous3fig1}b that collective anisotropy is larger than individual anisotropy at high density. Altogether, Fig.\ \ref{fig:autologous3fig2}b confirms that our findings at the level of anisotropy (Fig.\ \ref{fig:autologous3fig1}b) are also borne out at the level of migration.

\section{Discussion}

We have demonstrated that collective effects allow cells at high density to detect fluid flow and migrate downstream using autologous chemotaxis. Indeed, using theory, numerics, and simulation, we have shown that whereas individual autologous chemotaxis worsens with cell density, collective autologous chemotaxis improves. We derived and validated the associated scaling laws and identified a crossover cell density at which the optimal strategy switches from individual to collective. We observed this crossover in motility simulations invoking only cell-cell repulsion, and concentration and gradient sensing by single cells.

Collective effects are ubiquitous in cell biology, and previous work has shown that they confer behaviors beyond those available to single cells. Collective effects can sharpen a cell behavior: in epithelial cells, collective sensing allows groups of cells to detect shallower gradients than any single cell can detect alone \cite{ellison2016cell}. Collective effects can reverse a behavior: in lymphocytes, single cells migrate down a gradient, whereas groups of cells migrate up \cite{malet2015collective}. Here, we have found that collective effects can ``rescue'' a behavior: as cell density increases, individual sensing fails, but then collective sensing takes over and ultimately surpasses individual sensing. This is a potentially new interplay between single-cell and collective sensing that may suggest a density-dependent switch between two sensory regimes.

Sensory computations in single cells are performed by biochemical networks. It is not obvious that analogous computations can be performed collectively by groups of cells, especially when those cells are separated in space. In principle, the components of such a biochemical computation would need to be relayed diffusively among cells \cite{ellison2016cell, mugler2016limits}. Surprisingly, here we have found that in the case of autologous chemotaxis, the sensed signal and the relay signal can be the same component. The secreted molecule drifts with the flow, and thus its concentration is the sensed signal. At the same time, the secreted molecule originates from the cells themselves, and thus its concentration contains information on the cells' configuration; it is the relay signal. Even beyond sensing, the secreted molecule aids in collective migration because it acts as the coattractant. These simultaneous capabilities prevent the need for complicated extracellular secretion networks. Indeed, for the particular task of flow sensing by autologous chemotaxis, our results demonstrate that collective chemotaxis can be achieved with a single molecular species, and with the simple ingredients of concentration sensing and cell-cell repulsion.

Collective migration has not been observed in experiments on autologous chemotaxis performed to date. Perhaps this is because autologous chemotaxis has been discovered exclusively in eukaryotic cells, which are generally thought to migrate by comparing concentrations in space, whereas our mechanism requires comparing concentrations in time. Alternatively, perhaps this is because at high cell densities, where collective effects would dominate, it has been shown that a separate mechanism takes over that reverses migration, at least in breast cancer cells \cite{polacheck2011interstitial}. Nevertheless, the mechanism we reveal here is not specific to eukaryotic cells. Smaller cells such as bacteria use temporal sensing to track gradients. For such cells, in the presence of a flow, we predict that secreting and sensing a molecule is sufficient to produce efficient, collective migration in the flow direction.

\begin{acknowledgments}
This work was supported by National Science Foundation Grant Nos.\ MCB-1936761 and PHY-1945018.
\end{acknowledgments}

\providecommand{\noopsort}[1]{}\providecommand{\singleletter}[1]{#1}%

\end{document}